\documentclass[twocolumn,twoside,10pt]{extarticle}
\usepackage{amssymb}
\usepackage{enumerate}
\usepackage{a4wide}
\usepackage[english]{babel}
\usepackage[cmex10]{amsmath}
\usepackage{amsthm}
\newtheorem{definition}{Definition}
\usepackage{graphicx,algorithm,algorithmic,url,hyperref,authblk}

\usepackage[footnotesize]{caption}

\DeclareCaptionOption{parskip}[]{} % disable "parskip" caption option

\begin{document}
%\firstpage{1}

\title{SPATA: A Seeding and Patching Algorithm for Hybrid Transcriptome Assembly}
\author{Tin Chi Nguyen\,$^{1}$, Zhiyu Zhao\,$^{2}$ and Dongxiao Zhu\,$^{1,*}$}
\affil{$^{1}$Department of Computer Science, Wayne State University, Detroit, MI 48202.\\
         $^{2}$ Children's Medical Center Research Institute, UT Southwestern Medical Center, Dallas, TX 75390.\\
         $^{*}$To whom correspondence should be addressed.}
%         \small{Email: tin.nguyenchi@wayne.edu, zhiyu.zhao@utsouthwestern.edu, dzhu@wayne.edu.}}
\date{}
\maketitle

\section*{Abstract}
Transcriptome assembly from RNA-Seq reads is an active area of bioinformatics research. The ever-declining cost and the increasing depth of RNA-Seq have provided unprecedented opportunities to better identify expressed transcripts. However, the nonlinear transcript structures and the ultra-high throughput of RNA-Seq reads pose significant algorithmic and computational challenges to the existing transcriptome assembly approaches, either reference-guided or \textit{de novo}. While reference-guided approaches offer good sensitivity, they rely on alignment results of the splice-aware aligners and are thus unsuitable for species with incomplete reference genomes. In contrast, \textit{de novo} approaches do not depend on the reference genome but face a computational daunting task derived from the complexity of the graph built for the whole transcriptome. In response to these challenges, we present a hybrid approach to exploit an incomplete reference genome without relying on splice-aware aligners. We have designed a split-and-align procedure to efficiently localize the reads to individual genomic loci, which is followed by an accurate \textit{de novo} assembly to assemble reads falling into each locus. Using extensive simulation data, we demonstrate a high accuracy and precision in transcriptome reconstruction by comparing to selected transcriptome assembly tools. Our method is implemented in assemblySAM, a GUI software freely available at \url{http://sammate.sourceforge.net}.

\section{Introduction}
Human transcriptomes are highly diverse, overlapping, complex, and dynamic. Alternative splicing and structural variations play important roles in enhancing the range of transcriptome complexity~\cite{MatlinArianne2005,MortazaviAli2008,WangEric2008,Maher2009,MedvedevPaul2009,MedvedevPaul2010,Wang2009,Barash2010,NguyenTin2012}. For example, it is reported that over $90\%$ of human genes are alternatively spliced, and up to $5\%$ of structural variations, such as insertions and deletions are present within exons~\cite{FeukLars2006,PanQun2008,CroftLarry2000}. Moreover, all these events are condition-specific that lead to diversity of human transcriptomes. Identifying the expressed transcript sequences is a central task in transcriptomics research since it provides critical information for further analysis of the transcriptome. The ever-declining cost and increasing depth of RNA-Seq provide unprecedented opportunities to better identify expressed transcripts~\cite{OzsolakFatih2010}. Current efforts to reconstruct the expressed transcripts from short RNA-Seq reads generally follow one of the two strategies: an \textit{ab initio} strategy, which is also called reference-guided, and a \textit{de novo} strategy~\cite{HaasBrian2010,MartinJeffrey2011,ManuelGarber2011}. Despite initial success, both strategies are facing numerous challenges as described below and there is still a need for new transcriptome assembly approaches.

The reference-guided transcriptome assembly programs, e.g., ERANGE~\cite{MortazaviAli2008}, G-Mo.R-Se~\cite{DenoeudFrance2008}, Scripture~\cite{GuttmanMitchell2010}, Cufflinks~\cite{Trapnell2010}, IsoInfer~\cite{FengJianxing2011}, CEM~\cite{LiWei2012}, and IsoLasso~\cite{LiWei2011}, generally follow three steps. First, RNA-Seq reads are mapped to the reference genome using splice-aware alignment tools, such as TopHat~\cite{TrapnellCole2009}, SplitSeek~\cite{AmeurAdam2010}, MapSplice~\cite{WangKai2010}, HMMSplicer~\cite{DimonMichelle2010}, SpliceMap~\cite{AuKin2010}, and ABMapper~\cite{LouShao-Ke2011}. Second, a connectivity or splice graph is built to represent all possible isoforms at each genomic locus. Finally, alternatively spliced isoforms are identified by considering different possible paths in the graph, using the original read sequences and the paired-end information to filter out unlikely isoforms.

The reference-guided approaches offer good sensitivity, i.e., they are able to reconstruct transcripts with low abundance. However, since they rely on the alignment to the reference genome, their performance can be compromised by a number of factors. One major factor is that the reference genome being used maybe incomplete and may differ for each individual due to ubiquitous gene structural variations, such as point mutations, insertions, deletions, and gene fusions~\cite{Maher2009,MedvedevPaul2009,Wang2009,Kinsella2011}. Moreover, errors and biases introduced by the splice-aware aligners are carried over to the assembled transcripts. For example, junction reads spanning large introns can be missed because some aligners restrict the intron length due to the computational complexity~\cite{TrapnellCole2009,WangKai2010,AuKin2010,MartinJeffrey2011}. Consequently, the assembly programs may fail to reconstruct the corresponding transcripts.

In contrast, the \textit{de novo} transcriptome assembly approaches generally assemble reads using \emph{de Bruijn} graph~\cite{deBruijn1946, GoodIrving1946}, whose nodes are all subsequences of length $k$ (k-mers) extracted from the short reads whereas two nodes are connected by an directed edge if they perfectly overlap by $k-1$ nucleotides. The graph is then refined to eliminate possible false branches, using the original read sequences, the paired-end information and the coverage levels. The remaining paths are then traversed and reported as transcripts. This elegant computational solution was first introduced for whole genome assembly~\cite{PevznerPavel2001,ZerbinoDaniel2008,ButlerJonathan2008,SimpsonJared2009,PellJason2012}, where DNA sequencing depth is expected to be the same across the genome.

The \textit{de novo} transcriptome assembly programs, such as Rnnotator~\cite{MartinJeffrey2010}, Trans-Abyss~\cite{BirolInanc2009, RobertsonGordon2010}, Multiple-k~\cite{Surget-GrobaYann2010}, Trinity~\cite{Grabherr2011}, and Oases~\cite{SchulzMarcel2012}, don't depend on the reference genome and are able to reconstruct transcripts originated from genomes that have undergo rearrangements. However, a straightforward application of \emph{de Bruijn} graph on transcriptome assembly faces several challenges. First, the coverage depth fluctuates among transcripts, complicating the task of optimizing the trade-off between sensitivity and graph complexity~\cite{MartinJeffrey2011,ManuelGarber2011}. Second, the computing resources required to \textit{de novo} assemble the large transcriptomes as a whole may be overwhelming. Unlike the reference-guided transcriptome assembly programs, the \textit{de novo} transcriptome assembly programs do not use a divide-and-conquer strategy to divide the large assembly problem into many smaller assembly problems.

Here we present SPATA, a hybrid approach that combines the strengths of the previously mentioned strategies while avoiding their pitfalls. Comparing with the existing approaches, we claim the following original contributions: (1) We localize as opposed to align reads to individual genomic loci. (2) We develop a novel seeding and patching algorithm to \textit{de novo} assemble read sequences falling into each genomic locus. (3) We implement the multi-platform software assemblySAM, which takes RNA-Seq data in the FASTA or FASTQ formats as input and outputs transcript sequences.

\section{Methods}
SPATA is a novel hybrid approach that seeks to strike a good balance between the \textit{de novo} and the reference-guided transcriptome assembly approaches. On one hand, the difference between SPATA and the \textit{de novo} assembly approaches is that SPATA uses the existing reference genome to divide the large assembly problem into many smaller assembly problems and solve them independently. Since the graphs constructed from reads for each genomic locus are much less complex than that of the whole transcriptome, our approach promises a better performance and demands less computational resources. On the other hand, the difference between SPATA and the reference-guided assembly approaches is that SPATA does not rely on the exact mapping of the reads to the reference genome. While the reference guided approaches assemble the transcripts from a graph built from the mapped genomic positions, SPATA assembles read sequences falling into each genomic locus via a novel assembly algorithm.

Our approach proceeds into the following three stages: (1) Localization stage: we first align exonic reads to the reference genome. The remaining reads are then split and partially aligned to the reference genome. Reads that are aligned or partially aligned to a genomic locus are then used as input of the second stage. (2) Seeding and growing stage: we fully grow the backbone sequences starting from arbitrarily selected ``seeds''. The fully grown sequences are then used as input of the third stage. (3) Patching and cutting stage: we connect the backbone sequences to form an isoform graph and then traverse the paths of the graph to report the possible transcript sequences.

\subsection{Stage I: Localization}
\begin{figure}
\centering
\includegraphics[width=0.49\textwidth]{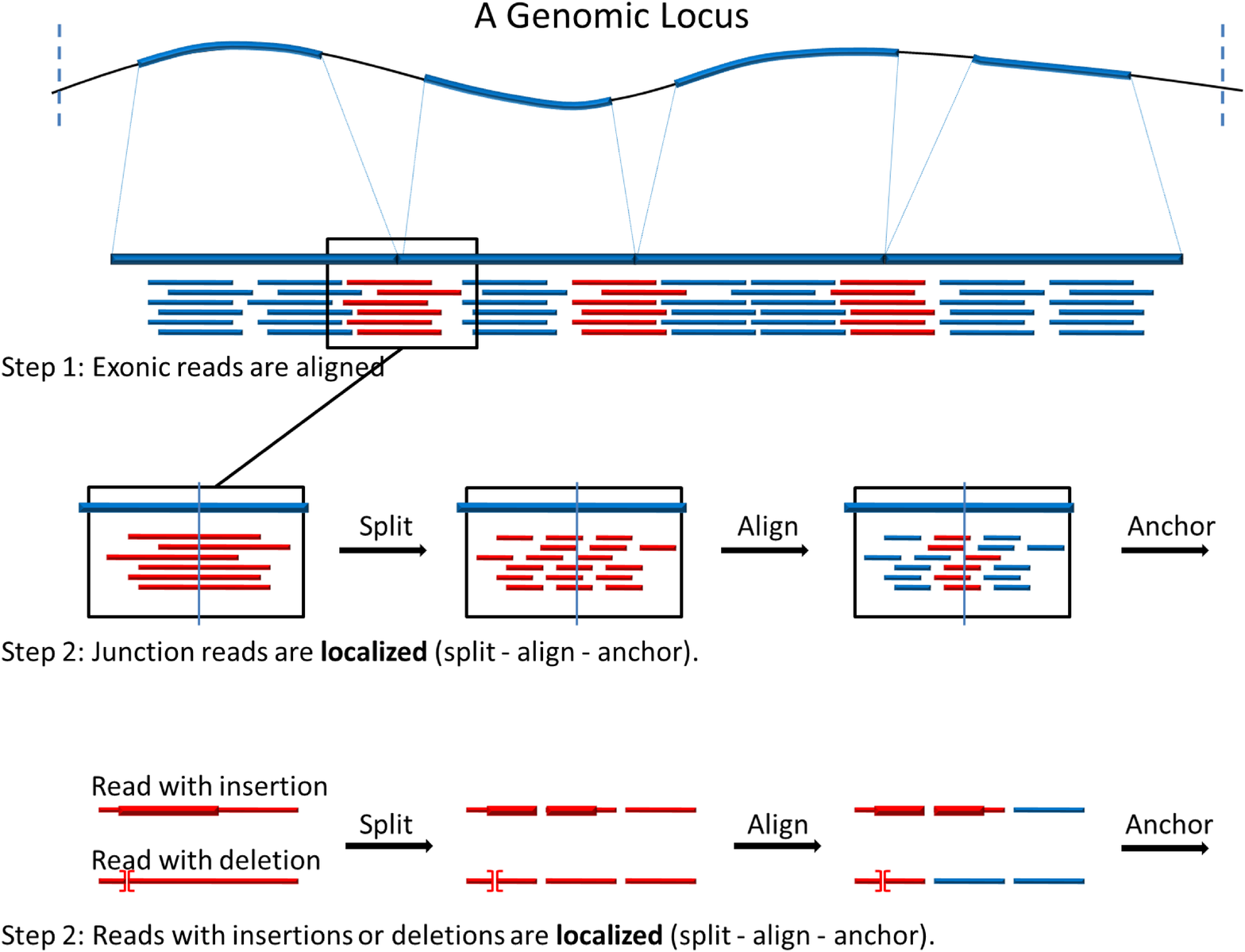}
\caption{Stage I: Localization. In \emph{step $1$}, RNA-Seq reads are aligned to the reference genome using a genome aligner. In this step, exonic reads are expected to be mapped to the reference genome. The unaligned reads are typically either junction reads or reads with insertions or deletions. In \emph{step $2$}, the remaining unaligned reads are split into three parts and then the split parts are aligned to the reference genome using a genome aligner. If any part of a split read is aligned, then the part is used as an anchor to localize the whole read to the genomic locus.}
\label{localization}
\end{figure}

We design a \emph{split-align-anchor} procedure to localize reads into genomic loci. We proceed into two steps as shown in Figure~\ref{localization}. Step $1$: Initial alignment of all the reads. We first align the entire set of reads, single-end or paired-end, to the reference genome using a genome aligner, e.g., Bowtie $2$~\cite{LangmeadBen2009,LangmeadBen2012}. For single-end reads, this step initially aligns most of the human exonic reads to the reference genome. For paired-end reads, we localize the reads if at least one read of the pair is aligned to the reference genome, including reads with structural variations. Step $2$: Follow-up localization of the split reads. For the remaining reads from step $1$, if they are single-end, we equally split each read into three partitions and localize the split reads using anchors. Anchors refer to any partition read(s) that are aligned to the reference genome. For paired-end reads, we equally split each read into two partitions, four partitions for a read pair, and localized them using anchors.

The main difference between our localization procedure and splice-aware aligners is that our procedure does not attemp to map every single nucleotide to the reference genome, which can be computational intensive or even impossible for reads with insertions or deletions. Instead, our goal is to localize reads to genomic loci. Reads falling into each genomic locus will serve as input of our \textit{de novo} assembly algorithm, which will be described in sections~\ref{sec:seeding} and ~\ref{sec:patching}. Below is the mathematical description of our localization procedure.

\subsubsection{Definitions}
\begin{definition}[Short read]
A short read $R$ is a sequence of $l$ letters, $(r_1 r_2 ... r_l)$, where $r_i \in \{A, T, C, G\}$ for $1 \le i \le l$. For convenience of the algorithm description, we denote any sequence of letters taken from $\{A, T, C, G\}$ in the same manner. We denote the length of a sequence as $|R|$. We also denote an empty sequence as $()$. A paired-end read of two short reads $R_1$ and $R_2$  is denoted as $(R_1,R_2)$.
\end{definition}

\begin{definition}[Split]
Given a short read $R$, we define the function $Split(R,2)$ that splits the read into $2$ subsequences $R_1$ and $R_2$ of approximately equal lengths. If $|R|$ is an even number, then $|R_1|=|R_2|=\frac{|R|}{2}$. Otherwise $|R_1|=\frac{|R|+1}{2}$ and $|R_2|=|R|-|R_1|$. Similarly, we define the function $Split(R,3)$ that splits the read $R$ into $3$ subsequences of approximately equal lengths.
\end{definition}

\subsubsection{The Reads Localization Algorithms}

\vspace{.2in}
\noindent {\bf Algorithm} Reads Localization and Binning (single-end)

\begin{algorithmic}
\REQUIRE A set of single-end reads $R$, a predefined intergenic distance $d$
\ENSURE Binning results $L$ \COMMENT{$L$ is a set of bins, each bin consists of reads falling into a genomic locus}\\
\vspace{.1in}
\COMMENT{Align exonic reads}
\STATE $P \leftarrow \emptyset$ \COMMENT{Initialize the localization result set}
\STATE Align $R$ to the reference genome using a genome aligner \COMMENT{Default is Bowtie $2$}
\FORALL {read $R_i \in R$}
\IF{$R_i$ \textit{is aligned to the reference genome at position} $P_i$}
\STATE $P \leftarrow P \cup \{(R_i,P_i)\}$
\STATE $R \leftarrow R - \{R_i\}$
\ENDIF
\ENDFOR\\
\vspace{.1in}
\COMMENT{Split and localize the rest}
\STATE $S \leftarrow \emptyset$ \COMMENT{Initialize the split set}
\FORALL { leftover read $R_i \in R$}
\STATE $(R_{i1},R_{i2},R_{i3}) \leftarrow Split(R_i,3)$ \COMMENT{Split into $3$}
\STATE $S \leftarrow S \cup \{(R_{i1},R_{i2},R_{i3})\}$
\ENDFOR
\STATE Align S to the reference genome using a genome aligner \COMMENT{Default is Bowtie $2$}
\FORALL {triple $(R_{i1},R_{i2},R_{i3}) \in S$}
\IF{$R_{i1}, R_{i2}$, \textit{or} $R_{i3}$ \textit{is aligned to the reference genome at position} $P_i$}
\STATE $P \leftarrow P \cup \{(R_i,P_i)\}$
\ENDIF
\ENDFOR\\
\vspace{.1in}
\COMMENT{Bin the reads to genomic loci}
\STATE Sort $P$ by increasing order of genomic position.
\STATE $L \leftarrow \emptyset$ \COMMENT{Initialize the binning sets}
\STATE $B \leftarrow \{(R_1,P_1)\}$ \COMMENT{Initialize the current bin as the read with smallest genomic position}
\FORALL {read-position pair $(R_i,P_i) \in S$ and $i>1$}
\IF{$P_i - P_{i-1} > d$}
\STATE $L \leftarrow L \cup \{B\}$ \COMMENT{Add the new bin}
\STATE $B \leftarrow \{(R_i,P_i)\}$ \COMMENT{Reset the bin}
\ELSE
\STATE $B \leftarrow B \cup \{(R_i,P_i)\}$ \COMMENT{Add new read to the current bin}
\ENDIF
\ENDFOR
\STATE $L \leftarrow L \cup \{B\}$ \COMMENT{Add the last bin}
\end{algorithmic}

\vspace{.2in}
\noindent {\bf Algorithm} Reads Localization and Binning (paired-end)

\begin{algorithmic}
\REQUIRE A set of paired-end reads $R$, a predefined intergenic distance $d$
\ENSURE Binning results $L$ \COMMENT{$L$ is a set of bins, each bin consists of reads falling into a genomic locus}\\
\vspace{.1in}
\COMMENT{Align exonic reads}
\STATE $P \leftarrow \emptyset$ \COMMENT{Initialize the localization result set}
\STATE Align $R$ to the reference genome using a genome aligner \COMMENT{Default is Bowtie $2$}
\FORALL {read $R_i=(R_{i1},R_{i2}) \in R$}
\IF{$R_{i1}$ or $R_{i2}$ \textit{is aligned to the reference genome at position} $P_i$}
\STATE $P \leftarrow P \cup \{(R_i,P_i)\}$
\STATE $R \leftarrow R - \{R_i\}$
\ENDIF
\ENDFOR\\
\vspace{.1in}
\COMMENT{Localize the rest}
\STATE $S \leftarrow \emptyset$ \COMMENT{Initialize split set}
\FORALL {leftover read $R_i=(R_{i1},R_{i2}) \in R$}
\STATE $(R_{i11},R_{i12}) \leftarrow Split(R_{i1},2)$ \COMMENT{Split into $2$}
\STATE $(R_{i21},R_{i22}) \leftarrow Split(R_{i2},2)$ \COMMENT{Split into $2$}
\STATE $S \leftarrow S \cup \{(R_{i11},R_{i12},R_{i21},R_{i22})\}$
\ENDFOR
\STATE Align $S$ to the reference genome using a genome aligner \COMMENT{Default is Bowtie $2$}
\FORALL {quadruple $(R_{i11},R_{i12},R_{i21},R_{i22}) \in S$}
\IF{$R_{i11},R_{i12},R_{i21},$ \textit{or} $R_{i22}$ \textit{is aligned to the reference genome at position} $P_i$}
\STATE $P \leftarrow P \cup \{(R_i, P_i)\}$
\ENDIF
\ENDFOR\\
\vspace{.1in}
\COMMENT{Bin the reads to genomic loci}
\STATE Sort $P$ by increasing order of genomic position.
\STATE $L \leftarrow \emptyset$ \COMMENT{Initialize the binning sets}
\STATE $B \leftarrow \{(R_1,P_1)\}$ \COMMENT{Initialize the current bin as the read with smallest genomic position}
\FORALL {read-position pair $(R_i,P_i) \in S$ and $i>1$}
\IF{$P_i - P_{i-1} > d$}
\STATE $L \leftarrow L \cup \{B\}$ \COMMENT{Add the new bin}
\STATE $B \leftarrow \{(R_i,P_i)\}$ \COMMENT{Reset the bin}
\ELSE
\STATE $B \leftarrow B \cup \{(R_i,P_i)\}$ \COMMENT{Add new read to the current bin}
\ENDIF
\ENDFOR
\STATE $L \leftarrow L \cup \{B\}$ \COMMENT{Add the last bin}
\end{algorithmic}

\subsection{Stage II: Seeding and Growing} \label{sec:seeding}
In the seeding and growing stage (stage II), our goal is to construct long backbone sequences from a set of short reads. At the beginning, all the reads are marked as ``active''. The algorithm then randomly picks a seed from the active set of reads and greedily extends it to both the left and right directions. If there are multiple reads that can extend a seed sequence, the read that most extends the seed sequence is chosen. After the sequence is fully grown, all the reads covered by the seed sequence are marked as inactive. For those remaining active reads, the algorithm repeatedly grows other backbone sequences until no active read left. To ensure that these fully grown backbone sequences are connected to each others, we extend each of them with one inactive read to both the left and right directions if possible. The extended ends will serve as connecting points to construct the isoform graph in the cutting and patching stage (stage III).

\begin{figure*}
\centering
\includegraphics[width=0.99\textwidth]{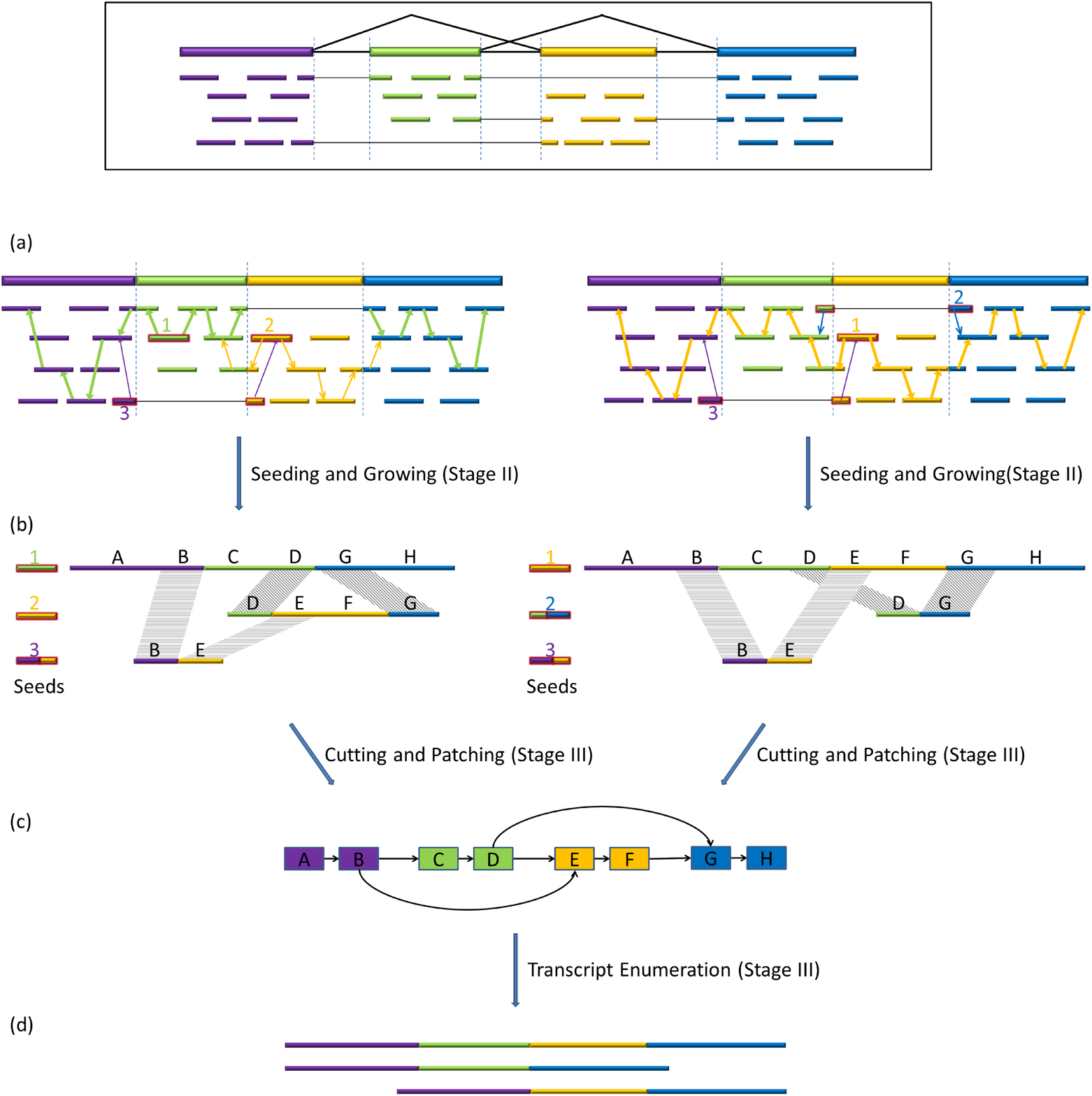}
\caption{Stages II and III: The transcriptome assembly algorithm. In this example, the RNA-Seq reads originated from three different isoforms: the first isoform consists of all the $4$ exons, the second isoform skips the yellow exon and the third isoform skips the green exon. The panel (a) shows the seeding and growing stage, where different choices of seeds may result in different backbone sequences. The arrows show the extension paths of the backbone sequences. In the \emph{left} panel, the first backbone sequence is grown from a green exonic seed and is extended to the leftmost purple read and to the rightmost blue read. The first backbone sequence spans across the purple, green, and blue exons. Likewise, the second sequence is grown from a yellow exonic read and the third sequence is grown from a purple-yellow junction read. The second backbone sequence covers the green-yellow junction, the yellow exon, and the yellow-blue junction whereas the third exon covers the purple-yellow junction. In the \emph{right} panel, the first sequence is grown from a yellow exonic seed and is extended to the leftmost purple read and to the rightmost blue read. The first backbone sequence spans across all $4$ exons. Likewise, the second sequence is grown from a green-blue junction read and the third sequence is grown from a purple-yellow junction read. The panel (b) displays the output backbone sequences of the seeding and growing stage, where the shadings show the overlaps between the backbone sequences. The seeding reads are displayed on the left of their grown sequences. The panel (c) displays the isoform graph after the cutting and patching stage, where the overlapping sequences serve as connecting points to glue the backbone sequences together. After this stage, the two sets of backbone sequences lead to the same isoform graph. The panel (d) displays the output transcript sequences.}
\label{fig:Seeding}
\end{figure*}

An example is displayed in Figure~\ref{fig:Seeding}a and b. In the seeding and growing stage, the seeds are randomly chosen. In the left panel of Figure~\ref{fig:Seeding}a, assuming that a green exonic read is chosen to be the first seed, the algorithm first extends the seed to the left until it reaches the leftmost purple read, then it extends the seed to the right until it reaches the rightmost blue read. After the first backbone sequence is constructed, the algorithm marks the following reads as inactive: purple exonic, purple-green junction, green exonic, green-blue junction, and blue exonic reads. The remaining active reads are purple-yellow junction, green-yellow junction, yellow exonic, and yellow-blue junction reads. Assuming a yellow exonic read is chosen to be the second seed, the algorithm grows the seed until it reaches the green-yellow junction read on the left and the yellow-blue junction read on the right. All active reads that are covered by this new sequence are marked as inactive. As described above, the second backbone sequence is then extended by one inactive green read on the left and by one inactive blue read on the right. After this step, only the purple-yellow junction read remains active and is chosen to be the third seed. Since there is no other active read, the seed is extended with one inactive purple read on the left and one inactive yellow read on the right. Similarly, in the right panel of Figure~\ref{fig:Seeding}a, the first seed is the highlighted yellow exonic read, the second seed is the green-blue junction read, and the third seed is the purple-yellow read. By choosing two different sets of seeds, the seeding and patching stage outputs two different sets of backbone sequences. However, after the cutting and patching stage, both sets of backbone sequences give rise to the same isoform graph as displayed in Figure~\ref{fig:Seeding}c, from which the same set of transcript sequences are enumerated. Below is the mathematical description of the seeding and growing algorithm.

\subsubsection{Definitions}
%\begin{definition}[Short read]
%A short read $R$ is a sequence of $l$ letters, $(r_1 r_2 ... r_l)$, where $r_i \in \{A, T, C, G\}$ for $1 \le i \le l$. For convenience of the algorithm description, we denote any sequence of letters taken from $\{A, T, C, G\}$ in the same manner. We also denote an empty sequence as $()$.
%\end{definition}

\begin{definition}[Overlap]
A short read $P=(p_1 p_2 ... p_l)$ overlaps with a short read $Q=(q_1 q_2 ... q_l)$ if $(p_{l-{l_o}+1} p_{l-{l_o}+2} ... p_l)=(q_1 q_2 ... q_{l_o})$ for ${l_o} \ge k$, where $l_o$ is the length of overlapped letters and $k$ is the minimum overlap cutoff. In this case we denote the overlapping length between the two short reads as $overlap(P, Q)=l_o$ and $overlap(Q, P)=-l_o$. If short reads $P$ and $Q$ do not overlap with each other, we denote $overlap(P, Q)=overlap(Q, P)=0$. Without loss of generality, we define the overlap between any two sequences in the same manner.
\end{definition}

\begin{definition}[Extension]
Given a short read $P$ and a set of short reads $T=\{T_1, T_2, ..., T_t\}$, a short read $T_r \in T$ is called a right extension of $P$ if $k \le overlap(P, T_r) \le overlap(P, T_i)$ for every $overlap(P, T_i) \ge k$ and $1 \le i \le t$. Similarly, $T_l \in T$ is called a left extension of $P$ if $k \le overlap(T_l, P) \le overlap(T_i, P)$ for every $overlap(T_i, P) \ge k$ and $1 \le i \le t$. We denote the right extension as $T_r=ext\_right(P, T)$ and the left extension as $T_l=ext\_left(P, T)$. We  also denote $ext\_left(P, T)=()$ or $ext\_right(P, T)=()$ if $P$ cannot be extended.
\end{definition}

\begin{definition}[Merger]
Two short reads can be merged to make a longer sequence if they overlap with each other. Given short reads $P=(p_1 p_2 ... p_l)$ and $Q=(q_1 q_2 ... q_l)$ with $overlap(P, Q)=l_o$, the merged sequence is $M=(p_1 p_2 ... p_{l-{l_o}} q_1 q_2 ... q_l)$ and is denoted as $M=merge(P, Q)$. Without loss of generality, we define the merger between any two sequences in the same manner.
\end{definition}

\begin{definition}[Cover and Subsequence]
Given two sequences $S=(s_1 s_2 ... s_l)$ and $S'=(s'_1 s'_2 ... s'_{l'})$, $S'$ is covered by $S$ if $l \ge l'$ and $(s_{i+1} s_{i+2} ... s_{i+l'})=(s'_1 s'_2 ... s'_{l'})$ for some $1 \le i \le l$. Denote it as $cover(S', S)=True$. We also say that $S'$ is a subsequence of $S$ and denote it as $S'=sub(S, i, l')$.
\end{definition}

\subsubsection{The Seeding and Growing Algorithm}
\vspace{.2in}
\noindent {\bf Algorithm} Seeding and Growing

\begin{algorithmic}
\REQUIRE A set of short reads $T=\{T_1, T_2, ..., T_t\}$ and the minimum overlap length $k$
\ENSURE A set of sequences $S$
\vspace{.1in}
\STATE $S \leftarrow \emptyset$ \COMMENT{Initialize $S$}

\WHILE {$T \ne \emptyset $}

\STATE $P \leftarrow T_1$   \COMMENT{Set $P$ as the first short read in $T$}
\STATE $S_P \leftarrow P$ \COMMENT{Initialize $S_P$ as the sequence of $P$}
\REPEAT
\STATE $T_l \leftarrow ext\_left(P, T)$
\STATE $S_P \leftarrow merge(T_l, S_P)$
\FORALL{$T_i \in T$ \AND $T_i \ne P$}
\IF{$cover(T_i, merge(T_l, P))=True$}
\STATE $T \leftarrow T - \{T_i\}$   \COMMENT{Remove $T_i$ from $T$}
\ENDIF
\ENDFOR
\STATE $P \leftarrow T_l$   \COMMENT{Set $P$ as its left extension}
\UNTIL{$ext\_left(P, T)=()$}

\STATE $P \leftarrow T_1$   \COMMENT{Set $P$ as the first short read in $T$}
\REPEAT
\STATE $T_r \leftarrow ext\_right(P, T)$
\STATE $S_P \leftarrow merge(S_P, T_r)$
\FORALL{$T_i \in T$}
\IF{$cover(T_i, merge(P, T_r))=True$}
\STATE $T \leftarrow T - \{T_i\}$   \COMMENT{Remove $T_i$ from $T$}
\ENDIF
\ENDFOR
\STATE $P \leftarrow T_r$   \COMMENT{Set $P$ as its right extension}
\UNTIL{$ext\_right(P, T)=()$}

\STATE $S \leftarrow S \cup \{S_P\}$    \COMMENT{Add sequence $S_P$ to set $S$}

\ENDWHILE
\end{algorithmic}

\subsection{Stage III: Patching and Cutting} \label{sec:patching}
\begin{figure*}
\centering
\includegraphics[width=0.99\textwidth]{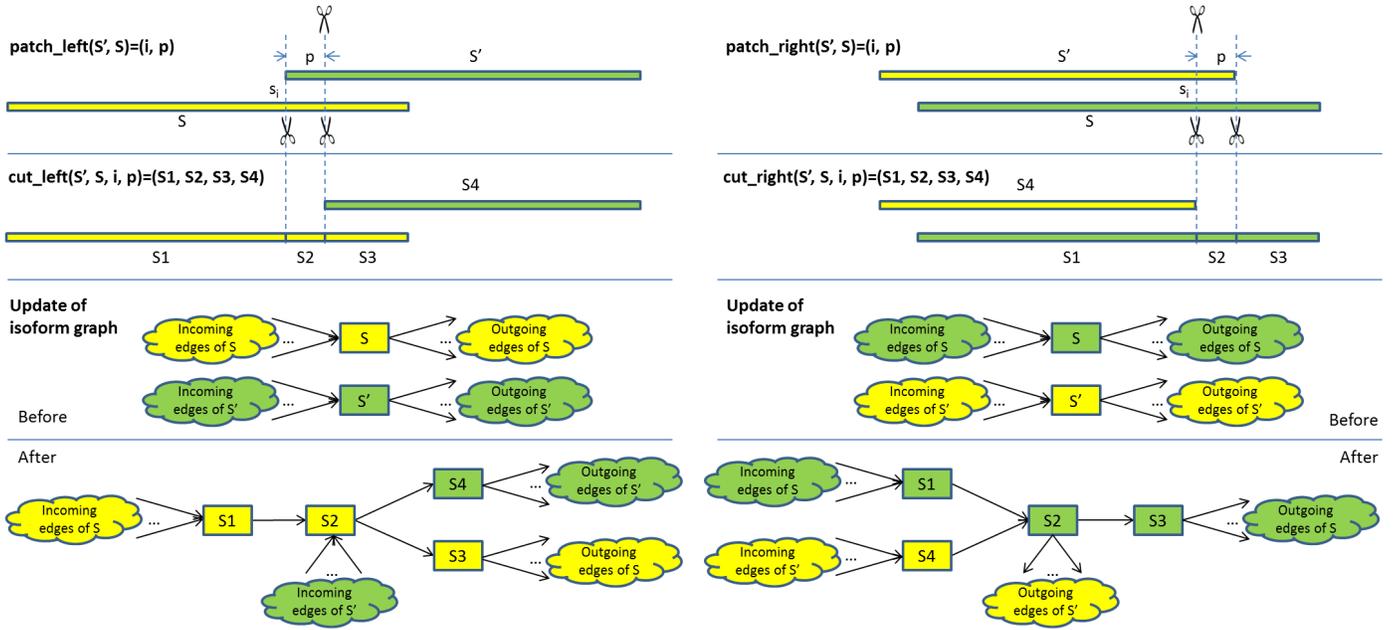}
\caption{A closer look at the patching and cutting stage (stage III). The \emph{left} panel displays the change of the isoform graph when S' overlaps with S whereas the \emph{right} panel displays the change of the isoform graph when the right end of S' overlaps with S. In both cases, the algorithm creates $4$ new vertices S1, S2, S3, and S4 and updates the edges accordingly before deleting S and S'.}
\label{fig:Patching}
\end{figure*}

Given a set of sequences outputted by the seeding and growing algorithm (stage II), the patching and cutting algorithm (stage III) dynamically constructs an isoform graph. Initially, vertices in the graph are set as sequences from the input and the connection relationship (connected or not connected) between them is ``unknown''. For every pair of vertices, the algorithm checks their connection relationship flag. If it is ``unknown'', the two vertices are tested for patching. If patched, then their sequences are cut into segments, the vertices are split, directed edges are added, and the relationship flags between all involved vertices are updated accordingly. Sometimes this patching and cutting process may cause unnecessary cuts, i.e., it may generate some pairs of vertices connected only by one edge in the graph. If this happens, the two vertices are merged into one. After the isoform graph has been constructed, each linear path starting from a vertex with no incoming edges and ending at a vertex with no outgoing edge is put into an isoform structure set, and the joint sequence based on that path is put into a contigs set (Figure~\ref{fig:Patching}). Below is the mathematical description of the patching and cutting algorithm.

\subsubsection{Definitions}
\begin{definition}[Patch]
Given two sequences $S=(s_1 s_2 ... s_l)$ and $S'=(s'_1 s'_2 ... s'_{l'})$, $S'$ is called a left patch of $S$ if we can find an $i$ and a $p \ge k$ for which $sub(S', 1, p)=sub(S, i, p)$. We denote it as $patch\_left(S', S)=(i, p)$. Similarly, $S'$ is called a right patch of $S$ if we can find an $i$ and a $p \ge k$ for which $sub(S', l'-p+1, p)=sub(S, i, p)$. We denote it as $patch\_right(S', S)=(i, p)$.
\end{definition}

\begin{definition}[Cut]
Given two sequences $S=(s_1 s_2 ... s_l)$ and $S'=(s'_1 s'_2 ... s'_{l'})$, if $patch\_left(S', S)=(i, p)$, a left cut between $S$ and $S'$ splits them into four subsequences $(s_1 s_2 ... s_{i-1})$, $(s_i s_{i+1} ... s_{i+p-1})$, $(s_{i+p} s_{i+p+1} ... s_l)$ and $(s'_{p+1} s'_{p+2} ... s'_{l'})$. We denote this as $cut\_left(S', S, i, p)=(S_1, S_2, S_3, S_4)$. Similarly, if $patch\_right(S', S)=(i, p)$, a right cut between $S$ and $S'$ splits them into four subsequences $(s_1 s_2 ... s_{i-1})$, $(s_i s_{i+1} ... s_{i+p-1})$, $(s_{i+p} s_{i+p+1} ... s_l)$ and $(s'_1 s'_2 ... s'_{l'-p})$. We denote this as $cut\_right(S', S, i, p)=(S_1, S_2, S_3, S_4)$.
\end{definition}

\begin{definition}[Joint]
Given two sequences $S=(s_1 s_2 ... s_l)$ and $S'=(s'_1 s'_2 ... s'_{l'})$, a joint sequence between them is $(s_1 s_2 ... s_l s'_1 s'_2 ... s'_{l'})$ and is denoted as $joint(S, S')$.  We also define the joint between $n$ sequences, $S_1, S_2, ..., S_n$ as $joint(S_1, S_2, ..., S_n)=joint(S_1, joint(S_2, joint\\(S_3, ..., joint(S_{n-1}, S_n))))$.
\end{definition}

\begin{definition}[Isoform graph, isoform structure and contig]
An isoform graph $G=(V, E)$ is a directed graph where each vertex is a sequence of letters from $\{A, T, C, G\}$ and each edge from vertex $V_i$ to vertex $V_j$ indicates that there exists a short read $R$ which is a subsequence of the joint sequence of $V_i$ and $V_j$, i.e., $cover(R, joint(V_i, V_j))=True$. Given an isoform graph $G=(V, E)$, an isoform structure is a linear path in $G$ starting at a vertex with no incoming edge and ending at a vertex with no outgoing edge. For an isoform structure with $n$ vertices as $V_1 \rightarrow V_2 \rightarrow ... \rightarrow V_n$, the corresponding contig is $joint(V_1, V_2, ..., V_n)$.
\end{definition}

\subsubsection{A Patching and Cutting Algorithm}
\vspace{.2in}
\noindent {\bf Algorithm} Patching and Cutting

\begin{algorithmic}
\REQUIRE A set of sequences $S$ and the minimum overlap length $k$
\ENSURE An isoform graph $G=(V, E)$, a set of isoform structures $I$, and a set of contigs $C$
\vspace{.1in}
\STATE $V \leftarrow S$ \COMMENT{Initialize $V$, a set of vertices}
\STATE $E \leftarrow \emptyset$ \COMMENT{Initialize $E$, a set of edges}
\FORALL{pair $V_i, V_j \in V$}
\STATE $F(V_i, V_j), F(V_j, V_i) \leftarrow ``unknown"$   \COMMENT{Initialize $F$, the relationship flags between vertices}
\ENDFOR
\FORALL{pair $V_i, V_j \in V$ \AND $F(V_i, V_j) = ``unknown"$}
\IF{$patch\_left(V_j, V_i)=(v, p)$}
\STATE $Update\_Graph(G, F, V_i, V_j, v, p, ``left")$
\ELSIF{$patch\_right(V_j, V_i)=(v, p)$}
\STATE $Update\_Graph(G, F, V_i, V_j, v, p, ``right")$
\ENDIF
\ENDFOR
\STATE $G \leftarrow Merge\_Vertices(G)$
\STATE $I \leftarrow Get\_Isoform\_Structures(G)$
\STATE $C \leftarrow \emptyset$ \COMMENT{Initialize $C$, a set of contigs}
\FORALL{isoform structure $I_i=V_1 \rightarrow V_2 \rightarrow ... \rightarrow V_n \in I$}
\STATE $C=C \cup \{joint(V_1, V_2, ..., V_n)\}$
\ENDFOR
\STATE
\end{algorithmic}

\vspace{.2in}
\noindent {\bf Procedure} Update\_Graph$(G, F, V_i, V_j, v, p, dir)$

\begin{algorithmic}
\REQUIRE Isoform graph $G=(V, E)$, flags variable $F$, vertices $V_i$ and $V_j$, cutting location variables $v$ and $p$, and direction variable $dir$
\ENSURE Updated graph $G$
\vspace{.1in}
\IF{$dir=``left"$}
\STATE $(V_1, V_2, V_3, V_4) \leftarrow cut\_left(V_j, V_i, v, p)$
\ELSE
\STATE $(V_1, V_2, V_3, V_4) \leftarrow cut\_right(V_j, V_i, v, p)$
\ENDIF
\STATE $V \leftarrow (V - \{V_i, V_j\}) \cup \{V_1, V_2, V_3, V_4\}$ \COMMENT{Remove two and add four vertices}
\FORALL{edge $V_m \rightarrow V_n \in E$}
\IF{$V_n=V_i$}
\STATE replace $V_n$ with $V_1$
\ENDIF
\IF{$V_m=V_i$}
\STATE replace $V_m$ with $V_3$
\ENDIF
\IF{$V_n=V_j$}
\IF{$dir=``left"$}
\STATE replace $V_n$ with $V_2$
\ELSE
\STATE replace $V_n$ with $V_4$
\ENDIF
\ENDIF
\IF{$V_m=V_j$}
\IF{$dir=``left"$}
\STATE replace $V_m$ with $V_4$
\ELSE
\STATE replace $V_m$ with $V_2$
\ENDIF
\ENDIF
\ENDFOR
\IF{$dir=``left"$}
\STATE $E \leftarrow E \cup \{V_1 \rightarrow V_2, V_2 \rightarrow V_3, V_2 \rightarrow V_4\}$ \COMMENT{Add three edges}
\ELSE
\STATE $E \leftarrow E \cup \{V_1 \rightarrow V_2, V_2 \rightarrow V_3, V_4 \rightarrow V_2\}$ \COMMENT{Add three edges}
\ENDIF
\STATE $Update\_Flags(F, V_i, V_j, V_1, V_2, V_3, V_4)$
\STATE
\end{algorithmic}

\vspace{.2in}
\noindent {\bf Procedure} Update\_Flags$(F, V_i, V_j, V_1, V_2, V_3, V_4)$

\begin{algorithmic}
\REQUIRE Flags variable $F$ and vertices $V_i, V_j, V_1, V_2, V_3, V_4$
\ENSURE Updated flags variable $F$
\vspace{.1in}
\STATE delete flags $F(V_i, V_i)$ and $F(V_j, V_j)$
\FORALL{relationship flag $F(V_m, V_n) \in F$}
\IF{$V_m=V_i$}
\STATE $F(V_1, V_n), F(V_2, V_n), F(V_3, V_n) \leftarrow F(V_m, V_n)$
\STATE $F(V_n, V_1), F(V_n, V_2), F(V_n, V_3) \leftarrow F(V_m, V_n)$
\STATE delete flags $F(V_m, V_n)$ and $F(V_n, V_m)$
\ENDIF
\IF{$V_m=V_j$}
\STATE $F(V_2, V_n), F(V_4, V_n) \leftarrow F(V_m, V_n)$
\STATE $F(V_n, V_2), F(V_n, V_4) \leftarrow F(V_m, V_n)$
\STATE delete flags $F(V_m, V_n)$ and $F(V_n, V_m)$
\ENDIF
\ENDFOR
\FOR{$1 \le m, n \le 4$}
\STATE $F(V_m, V_n) \leftarrow ``known"$
\ENDFOR
\STATE $F(V_3, V_4), F(V_4, V_3) \leftarrow ``unknown"$
\STATE
\end{algorithmic}

\vspace{.2in}
\noindent {\bf Procedure} Merge\_Vertices$(G)$

\begin{algorithmic}
\REQUIRE An isoform graph $G$
\ENSURE Graph $G$ with merged vertices
\vspace{.1in}
\FORALL{pair $V_i, V_j \in V$}
\IF{$V_i \rightarrow V_j \in E$ is the only connection between them}
\STATE $V_m \leftarrow joint(V_i, V_j)$ \COMMENT{Merge two vertices}
\STATE $V \leftarrow (V-\{V_i, V_j\}) \cup \{V_m\}$   \COMMENT{Remove two vertices and add the merged vertex}
\FORALL{$V_n \rightarrow V_i \in E$}
\STATE Replace $V_i$ with $V_m$ \COMMENT{Update incoming edges of $V_i$}
\ENDFOR
\FORALL{$V_j \rightarrow V_n \in E$}
\STATE Replace $V_j$ with $V_m$ \COMMENT{Update outgoing edges of $V_j$}
\ENDFOR
\ENDIF
\ENDFOR
\end{algorithmic}

\vspace{.2in}
\noindent {\bf Procedure} Get\_Isoform\_Structures$(G)$

\begin{algorithmic}
\REQUIRE An isoform graph $G$
\ENSURE An isoform structures set $I$
\vspace{.1in}
\STATE $I \leftarrow \emptyset$ \COMMENT{Initialize $I$}
\FORALL{vertex $V_i \in V$ that has no incoming edge}
\STATE $I \leftarrow \{V_i\}$
\ENDFOR
\FORALL{path $I_i=V_{i_1} \rightarrow V_{i_2} \rightarrow ... \rightarrow V_{i_n} \in I$}
\IF{vertex $V_{i_n}$ has outgoing edges}
\FORALL{edge $V_{i_n} \rightarrow V_{i_m} \in E$}
\STATE $I \leftarrow I \cup \{V_{i_1} \rightarrow V_{i_2} \rightarrow ... \rightarrow V_{i_n} \rightarrow V_{i_m}\}$
\ENDFOR
\STATE $I \leftarrow I - \{I_i\}$ \COMMENT{Remove $I_i$ from $I$}
\ENDIF
\ENDFOR
\end{algorithmic}

\subsection{Error control}
The above algorithms do not take into account the sequencing errors in short reads. In reality, due to those errors, the short reads do not always perfectly overlap with each other. Therefore the algorithms must have error tolerance capability in order to process real sequencing data. For this purpose, we slightly modify the following definitions:
\begin{definition}[Overlap, updated]
A short read $P=(p_1 p_2 ... p_l)$ overlaps with $Q=(q_1 q_2 ... q_l)$ if $(p_{l-{l_o}+1} p_{l-{l_o}+2} ... p_l)=(q_1 q_2 ... q_{l_o})$ for ${l_o} \ge k$ with at most $e_1 l_o$ differences (errors) and no more than $e_2$ contiguous errors, where $l_o$ is the length of overlapped letters, $k$ is the minimum overlap cutoff, $e_1$ is the maximum error rate in any short read, i.e., there are at most $e_1 * l$ sequencing errors in a short read of length $l$. Here $e_2$ is a small constant, for instance, $2$. Without loss of generality, we redefine the overlap between any two sequences in the same manner.
\end{definition}
\begin{definition}[Cover and Subsequence, updated]
Given two sequences $S=(s_1 s_2 ... s_l)$ and $S'=(s'_1 s'_2 ... s'_{l'})$, $S'$ is covered by $S$ if $l \ge l'$ and $(s_{i+1} s_{i+2} ... s_{i+l'})=(s'_1 s'_2 ... s'_{l'})$ for some $1 \le i \le l$ with at most $e_1 * l'$ differences (errors) and no more than $e_2$ contiguous errors. We also say that $S'$ is a subsequence of $S$.
\end{definition}
With these modified definitions, the proposed algorithms can be applied to process data with sequencing errors. %The algorithms need to take two parameters, $e_1$ and $e_2$, where $e_1$ is estimated by the user and $e_2$ is a fixed constant such as $2$.

\subsection{Complexity Analysis}

\subsubsection{The Localization and Binning Algorithm}
Given a read aligner with running time $T(g,l)$ per short read, where $g$ is the length of the reference genome and $l$ is the short read length, the localization stage runs in $O(T(g,l)r)$ time, where $r$ is the total number of short reads. Running time of the binning step is $O(r\log r + r)=O(r\log r)$ because sorting genomic position takes $O(r\log r)$ time and the binning time after sorting is $O(r)$. Therefore, the total time for localization and binning is $O(T(g,l)r + r\log r)$. Since $T(g,l)$ is $\Omega(g)$ and $g \gg \log r$, we have $O(T(g,l)r + r\log r)= O(T(g,l)r)$, i.e., the time complexity of the localization and binning algorithm is mainly determined by the time complexity of the genome aligner.

\subsubsection{The Seeding and Growing Algorithm}
There are three major operations in this algorithm: $ext()$, $merge()$, and $cover()$. Their time complexity depends on the basic $overlap()$ operation. For a short read with length $l$, the time complexity of the $overlap()$ operation is $O(l)$. Denoting the average number of short reads per bin as $m$, in each extension, operation $ext()$ runs in $O(ml)$ time because there are $m$ pairs of short reads to test for overlaps. The time complexity of $merge()$ is $O(l)$ because the merging locations are already known from the preceding $ext()$ operation. The time complexity of $cover()$ is also $O(l)$. Since the typical length of a grown sequence is $s$ and usually an $ext()$ operation extends a growing sequence by $O(l)$ letters, when seeding and growing a sequence, the number of times $ext()$ and $merge()$ operations are called is $O(\frac{s}{l})$. The number of $cover()$ operations, when its input is limited to those overlapped short reads discovered by the $ext()$ operation, is only $O(\frac{s}{l}o)$ where $o$ is the number of overlapped short reads discovered when extending a read. Therefore, the total time for growing one sequence is $O(\frac{s}{l}ml)+O(\frac{s}{l}l)+O(\frac{s}{l}ol)$=$O(sm)$ as $m>o$. Also, there are only a small number of such sequences to grow. Typically this number is $O(1)$. Thus the total time for growing all the sequences in one bin or transcriptional unit is $O(sm)$. The total time for growing all the sequences in $n$ bin or transcriptional units in a whole transcriptome is $O(smn)=O(sr)$ where r is the total number of short reads. Typical values of $s$ and $r$ are $O(10^3)$ and $O(10^7) \sim O(10^8)$, respectively.

\begin{figure}
\centering
\includegraphics[width=0.49\textwidth]{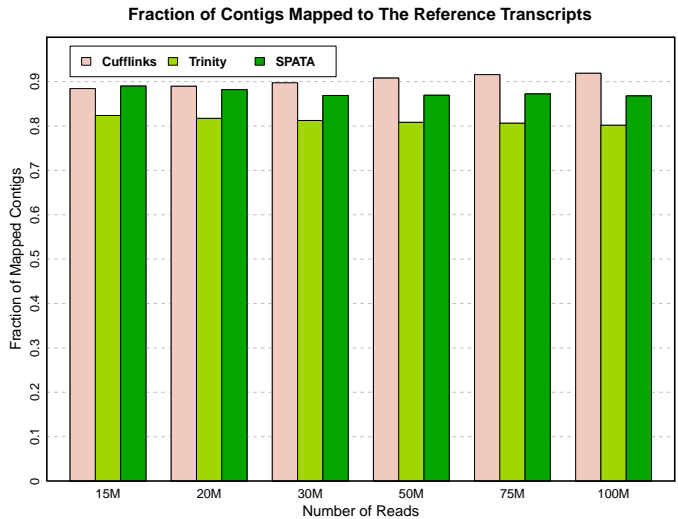}
\caption{Precision analysis of Cufflinks, Trinity, and SPATA using simulated data. The figure displays the fraction of contigs mapped to the reference transcripts. The horizontal axis displays the number of reads for each data set whereas the vertical axis displays the fraction of output contigs that can be mapped to the reference transcripts. A contig is considered to be mapped properly if at least $90\%$ of its sequence is covered by a reference transcript.}
\label{fig:Precision}
\end{figure}

\subsubsection{The Patching and Cutting Algorithm}
The time complexity of the patching and cutting algorithm is determined by the number of vertices and the number of edges in the isoform graph. Because there are only $O(1)$ sequences generated in the seeding and growing stage, typically the number of vertices in the final graph is $O(1) \sim O(10)$ and the number of edges is $O(1) \sim O(10^2)$ at most. As these numbers are much smaller than $s$, $m$, and $n$, the time complexity of the patching and cutting algorithm can be neglected.

The total time complexity of the algorithms for short reads localization, transcriptome assembly and isoforms reconstruction is thus $O(T(g,l)r+sr)$. Since $T(g,l)$ is $\Omega(g)$ and $g \gg s$, the complexity is mainly determined by the read mapping algorithm.

\section{Simulation Experiments}

\begin{figure*}
\centering
\includegraphics[width=0.49\textwidth]{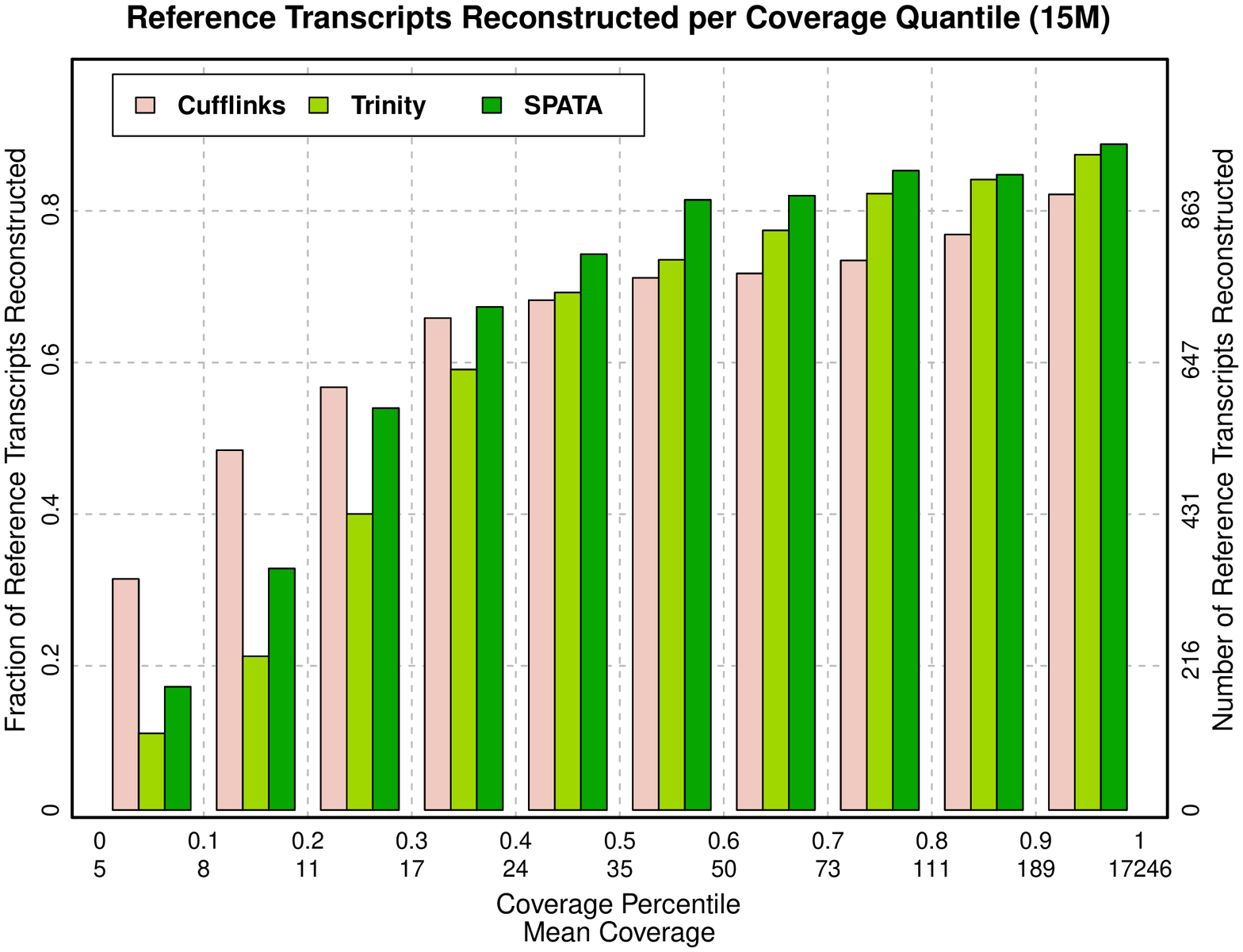}
\includegraphics[width=0.49\textwidth]{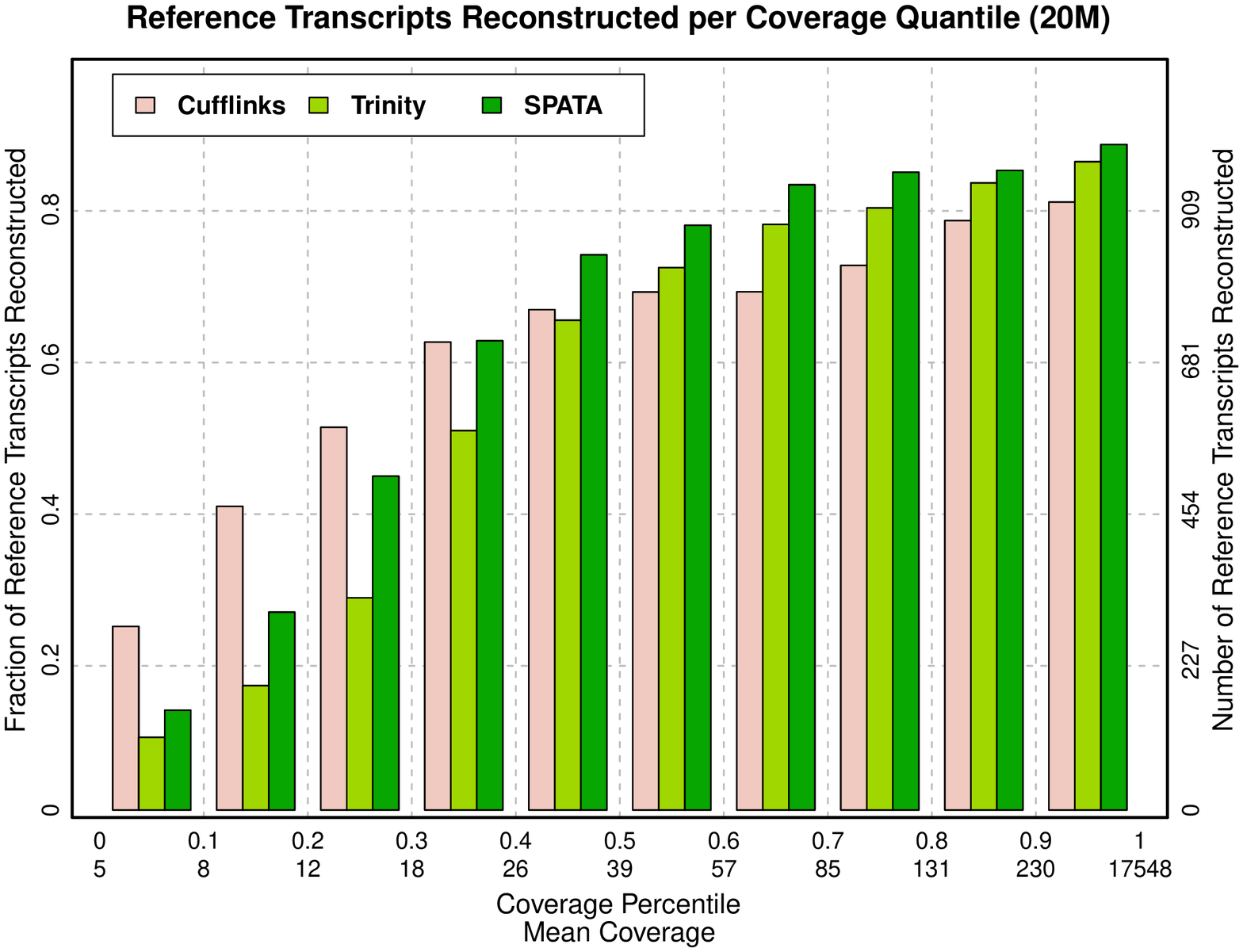}
\includegraphics[width=0.49\textwidth]{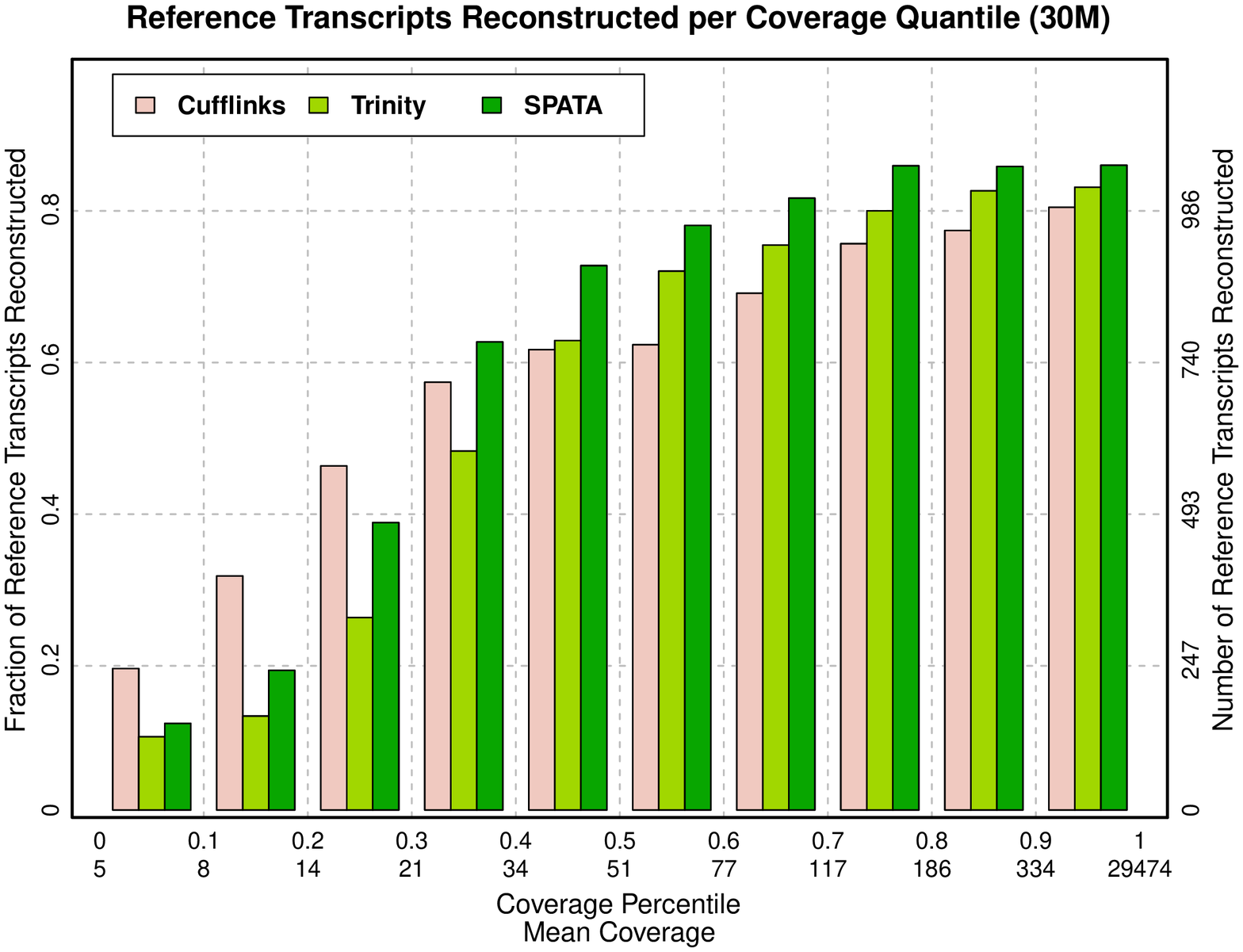}
\includegraphics[width=0.49\textwidth]{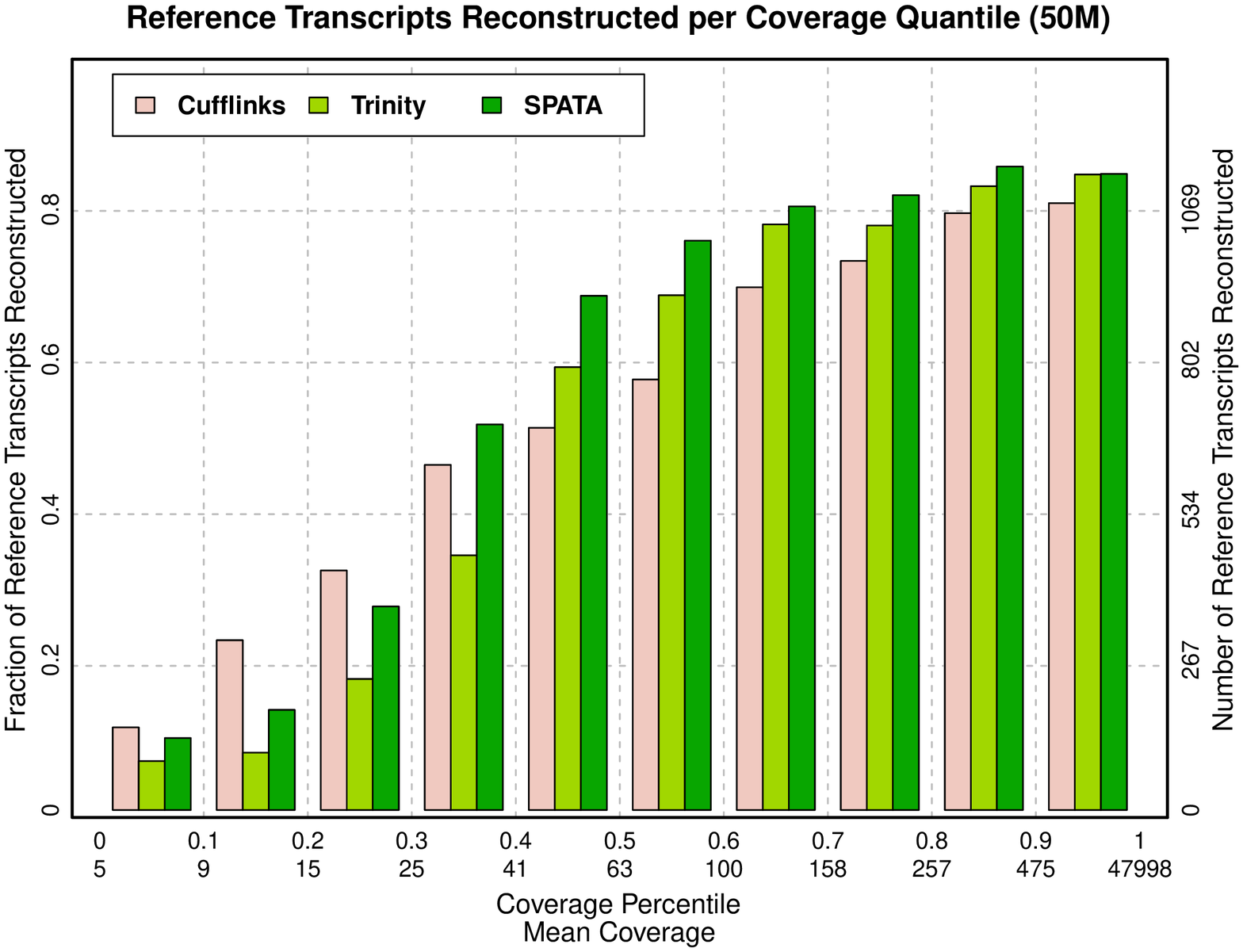}
\includegraphics[width=0.49\textwidth]{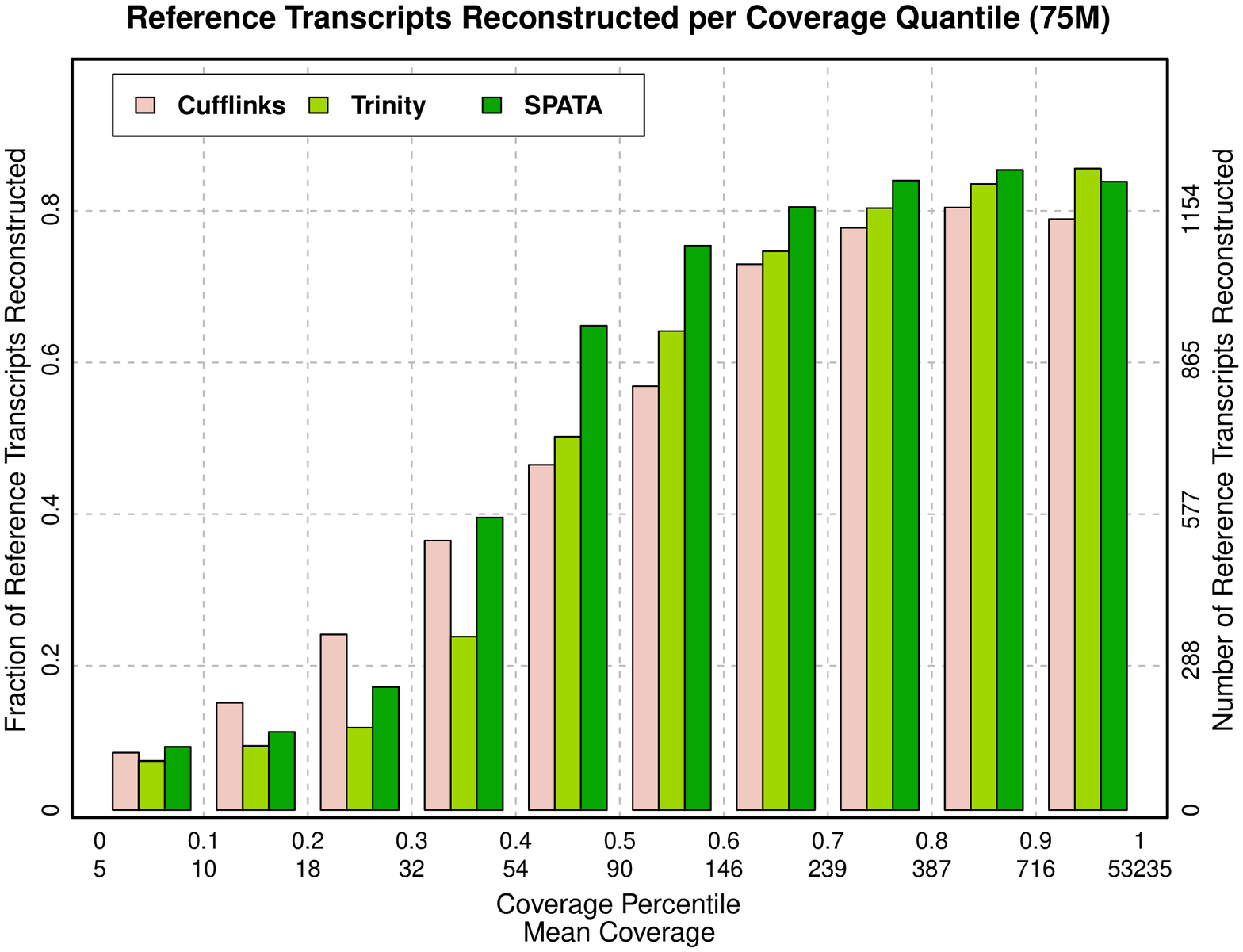}
\includegraphics[width=0.49\textwidth]{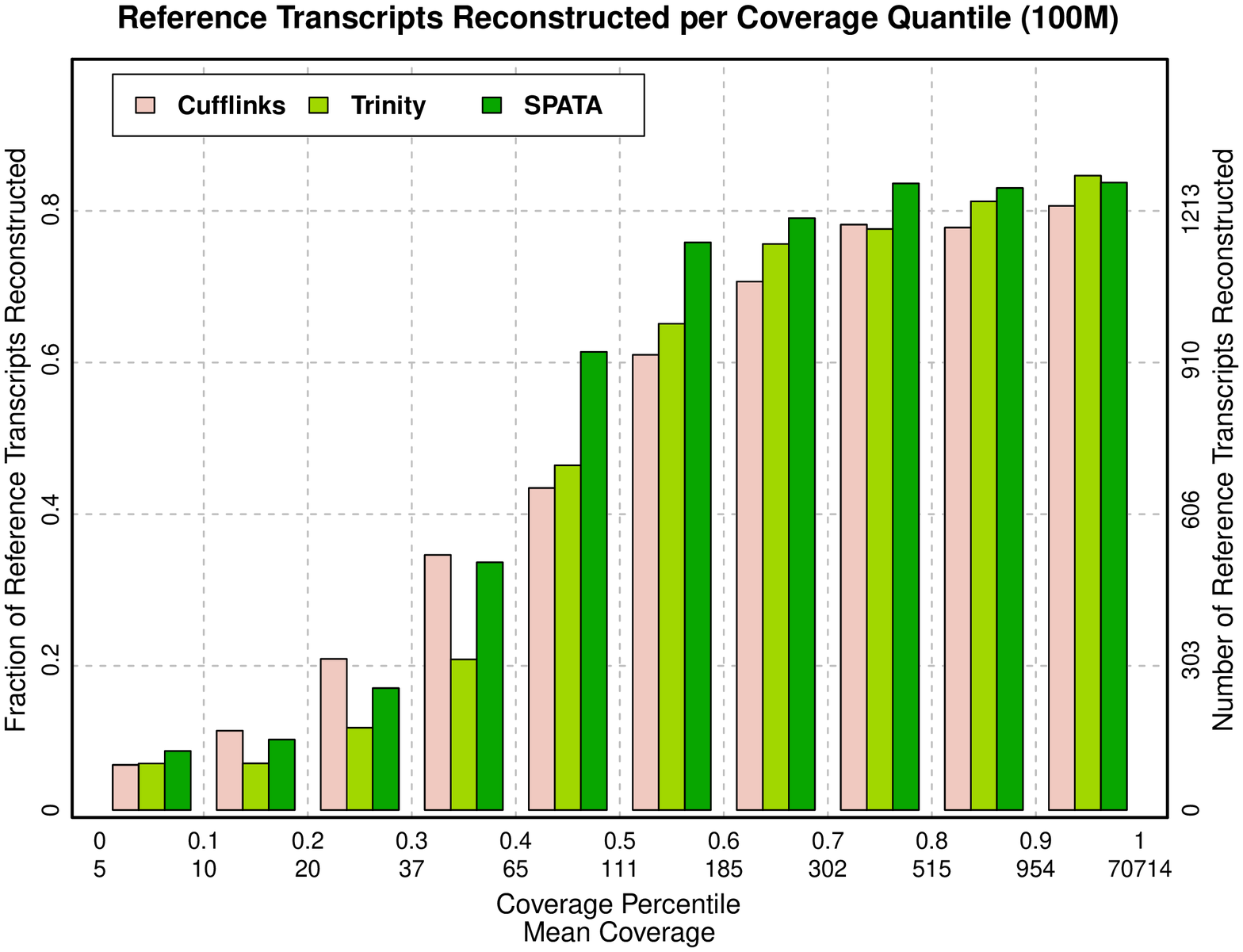}
\caption{Accuracy analysis of Cufflinks, Trinity, and SPATA using simulated data. The figure displays the fraction of reference transcripts fully reconstructed by different coverage quantiles. The six panels are arranged in an increasing order of read numbers from $15$ million to $100$ million of paired-end reads. From the ground truth, we know the reference transcripts and their expression abundances. In each panel, the horizontal axis shows the expression percentile (in $10\%$ increments) and the corresponding mean coverage of the reference transcripts. The left vertical axis shows the fraction of reference transcripts that can be fully reconstructed by each method. The right vertical axis displays the number of reference transcripts that were fully reconstructed by each method. A transcript is considered to be fully reconstructed by a method if it has at least $90\%$ base-identity with a contig from the method's output. We examine all reference transcripts having mean coverage of $5$ or higher.}
\label{fig:Sensitivity}
\end{figure*}

We used FluxSimulator [\url{http://flux.sammeth.net}], an open source software package simulating whole transcriptome sequencing experiments, to generate RNA-Seq data sets. FluxSimulator first randomly generates integer copies of each splicing isoform according to the user-provided annotation file. It then constructs an amplified, size-selected library and sequences the library \emph{in silico}. The resulting cDNA fragments are then sampled randomly for simulated sequencing, where the initial and terminal ends of each selected fragment are reported as reads. The current version of FluxSimulator (1.2) provides two sequencing error models, one is for read length $76$ and another is for read length $36$. The default error model was used to generate six data sets consisting of $15$ million, $20$ million, $30$ million, $50$ million, $75$ million, and $100$ million paired-end reads with length $76$ from the reference protein-coding transcripts available from the Ensembl database (version GRCh37.69). More details of the simulated data sets and the error model can be found in the Appendix. Both simulated RNA-Seq data sets and implementation of SPATA are available for download at \url{http://sammate.sourceforge.net/}.

We compared the performance of SPATA with Cufflinks~\cite{Trapnell2010} and Trinity~\cite{Grabherr2011}. Cufflinks is a reference-guided transcriptome assembly whereas Trinity is a \textit{de novo} transcriptome assembly~\cite{MartinJeffrey2011,ManuelGarber2011}. As mentioned before, SPATA is a hybridization of the reference-guided and the \textit{de novo} strategies, i.e., it uses the reference genome to localize reads to genomic loci and then assembles the read sequences falling into each genomic locus using a new assembly algorithm. Simulated FASTQ files were used as input for SPATA and Trinity. Since Cufflinks does not process raw reads, we used TopHat~\cite{TrapnellCole2009} to align reads from the FASTQ files to the human reference genome. We then used TopHat's BAM files as the input for Cufflinks. We used default parameters to run all the tools.

\subsection{Precision and Accuracy Analyses}

For each dataset, the ground truth is a set of expressed transcript sequences whereas the output of each assembly software is a set of contigs, i.e., a continuous sequence of bases constructed from the short reads in the FASTQ files. In the rest of the manuscript, we use the term \emph{reference transcripts} to refer to the expressed transcript sequences. Output contigs of each assembly program were compared against the reference transcripts using the pairwise sequence alignment program SSAHA2~\cite{Ning2001}. We evaluated and compared the performance of the assembly programs according to ``precision'' and ``accuracy''. When output contigs were aligned to the reference transcripts, we define the precision as the fraction of contigs that were properly mapped to the reference transcripts. A contig is considered to be mapped if at least $90\%$ of its sequence was covered by a reference transcript ($\frac{M}{M + N + G} \geq 90\%$, where $M$ is the number of matched bases, $N$ is the number of mismatched bases and $G$ is the cumulative insertion and deletion length). When the reference transcripts were aligned to the constructed contigs, we define the accuracy as the fraction of reference transcripts that can be fully reconstructed. A transcript is defined to be fully reconstructed if it has at least $90\%$ base-identity with one of the output contigs.

When evaluating the performance of different assembly software, high values for both precision and accuracy are important. On one hand, a high precision indicates that most of the method's contigs are subsequences of the reference transcripts. On the other hand, a high accuracy indicates that a method reconstructed a large number of the reference transcripts.

\begin{figure}
\centering
\includegraphics[width=0.49\textwidth]{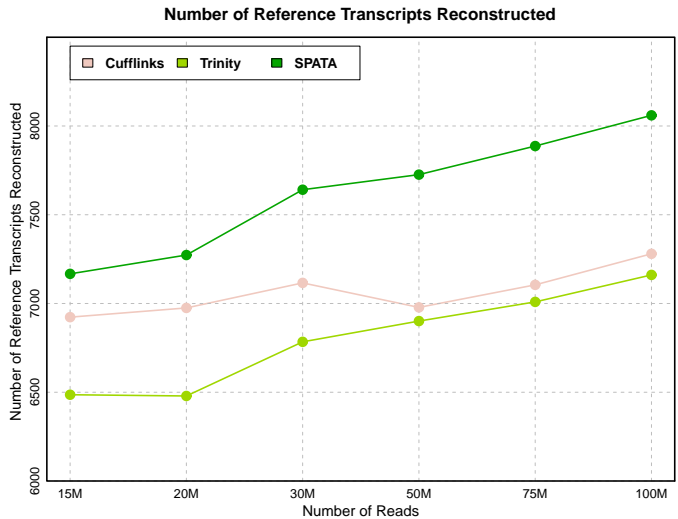}
\caption{The number of reference transcripts reconstructed by Cufflinks, Trinity, and SPATA using simulated data. The horizontal axis shows the number of reads for each data set in an increasing order whereas the vertical axis shows the total number of reference transcripts that were fully reconstructed. A transcript is consider to be fully reconstructed by a method if it has at least $90\%$ base-identity with a contig from the method's output. We examine all reference transcripts having mean coverage of $5$ or higher.}
\label{fig:Accuracy_All}
\end{figure}

Regarding precision, we calculated the fraction of output contigs of Cufflinks, Trinity, and SPATA that were properly mapped to the reference transcripts. Figure~\ref{fig:Precision} displays the fraction of contigs that were mapped to reference transcripts. The performance comparison in terms of precision is overall comparable with slightly higher values for Cufflinks. It might due to the following reasons: First, Cufflinks only processes reads that were aligned to the reference genome and discards reads with sequencing errors. Second, Cufflinks attempts to identify the minimum number of isoforms that cover the majority of reads.

\begin{figure*}
\centering
\includegraphics[width=0.49\textwidth]{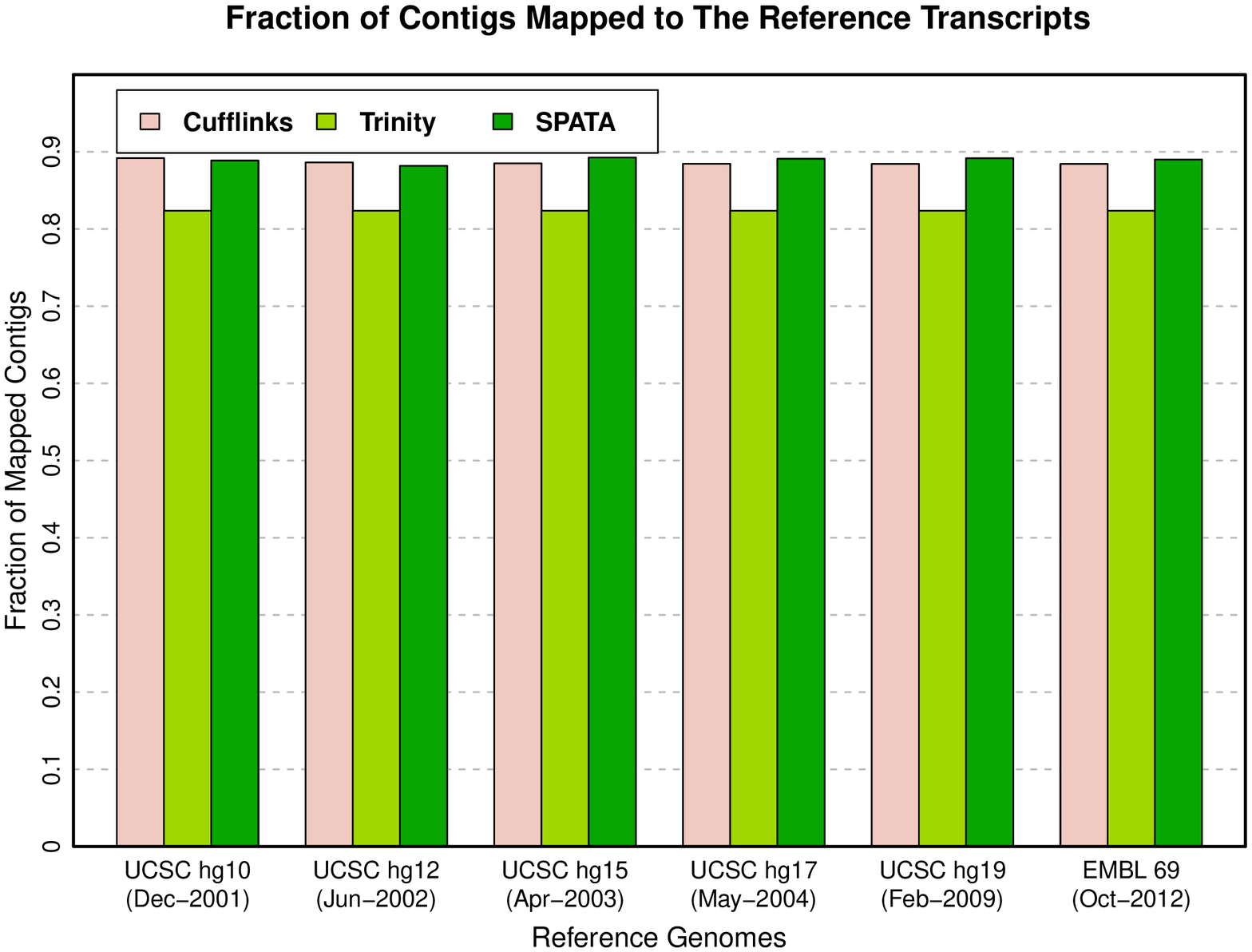}
\includegraphics[width=0.49\textwidth]{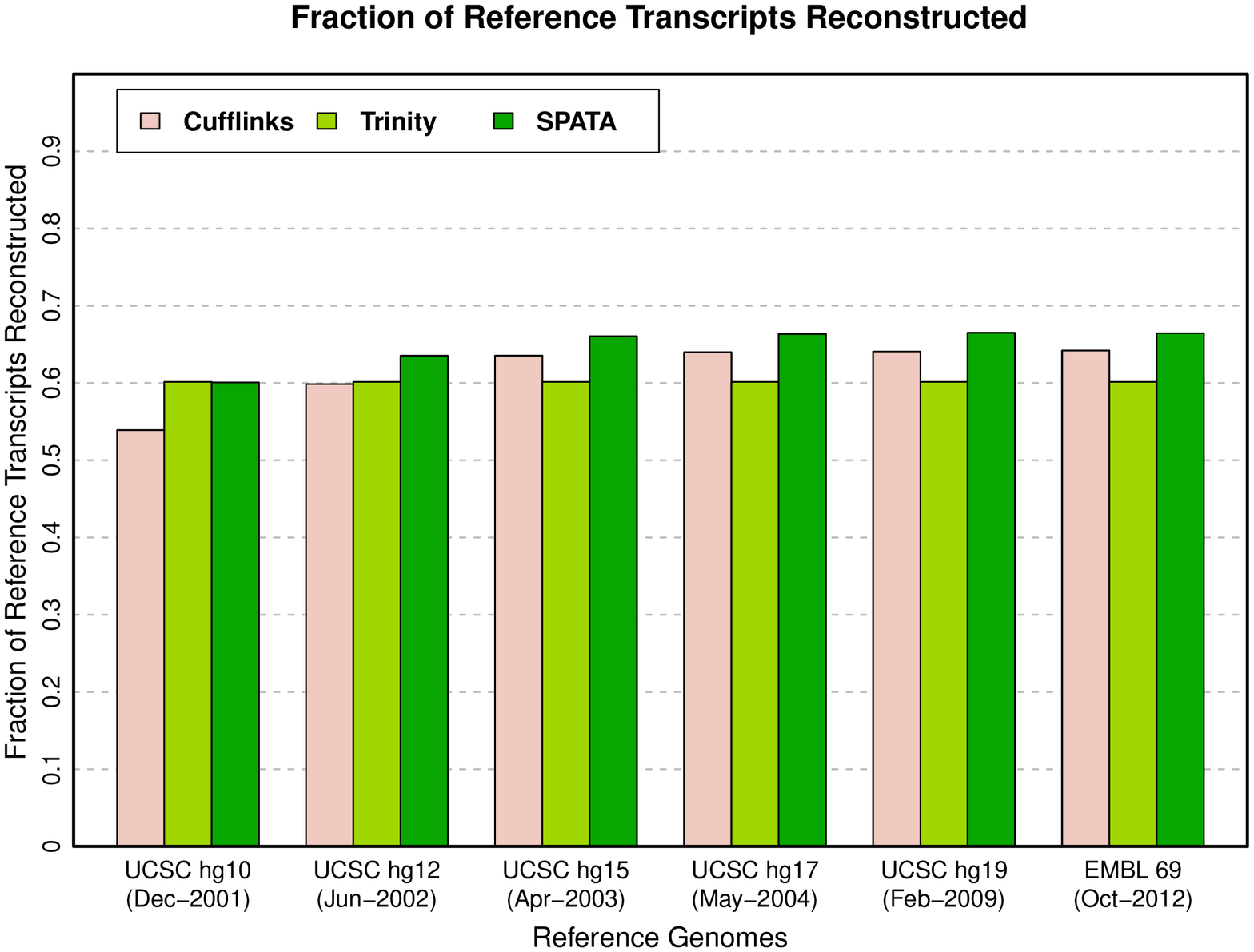}
\caption{Precision (\emph{left}) and accuracy (\emph{right}) analyses of Cufflinks, Trinity, and SPATA using different reference genomes for the data set consisting of $15$ million reads. The horizonal axes display the reference genomes being used in chronological order. The vertical axis in the \emph{left} panel displays the fraction of output contigs that can be mapped to the reference transcripts. A contig is considered to be mapped properly if at least $90\%$ of its sequence is covered by a reference transcript. The vertical axis in the \emph{right} panel displays the fraction of reference transcripts that can be fully reconstructed by each method. A transcript is consider to be fully reconstructed by a method if it has at least $90\%$ base-identity with a contig from the method's output. We examine all reference transcripts having mean coverage of $5$ or higher.}
\label{fig:Genomes_All}
\end{figure*}

Regarding accuracy, we calculated the fraction of reference transcripts that can be fully reconstructed by Cufflinks, Trinity, and SPATA. Figure~\ref{fig:Sensitivity} displays the number of reference transcripts that were fully reconstructed across a broad ranges of expression levels and sequencing depths. The expression of the reference transcripts is calculated from the simulated ground truth origin of the paired-end reads, i.e., the mean coverage of a transcript is calculated as the number of reads multiplied with the read length divided by the transcript length. We examined all reference transcripts having mean coverage of $5$ or higher.

In general, Cufflinks performed better at low expression levels whereas SPATA and Trinity performed better at high expression levels (Figure~\ref{fig:Sensitivity}). Cufflinks may perform better at lower expression levels since the small gaps within a transcript caused by sequencing errors or lack of read coverage can be corrected using the reference genome~\cite{Trapnell2010,MartinJeffrey2011,ManuelGarber2011}. Trinity and SPATA, however, outperform Cufflinks in assembling highly reference transcripts. A possible reason is that the parsimony strategy used by Cufflinks attempts to identify the minimum number of isoforms to cover the majority of reads while filtering out uncertain reads. The latter may result in missing some alternative splicing events~\cite{Trapnell2010,MartinJeffrey2011,ManuelGarber2011,LiWei2011}.

Another notable phenomenon is that SPATA reconstructs more reference transcripts than Cufflinks and Trinity for any of the six data sets. Figure~\ref{fig:Accuracy_All} displays the total number of reference transcripts reconstructed by each method across all of the six data sets. Interestingly, Trinity and SPATA reconstruct more reference transcripts as the number of reads increases whereas the number of reference transcripts reconstructed by Cufflinks fluctuates.

\subsection{Robust Analysis Using Incomplete Reference Genomes}

In the above analyses, Cufflinks and SPATA used the human reference genome version GRCh37.69 downloaded from Ensembl. The same genome sequence was used to generate the simulated data sets by FluxSimulator. We consider this latest version of human genome as the ``complete'' reference genome. In order to access the robustness of Cufflinks and SPATA against incomplete reference genomes, we also examined their performance using different versions of the human reference genome. We downloaded the following reference genome sequences from UCSC website: hg10 (December 2001), hg12 (June 2002), hg15 (April 2004), hg17 (February 2009), and hg19 (October 2012). The older a reference genome, the more incomplete it is compared to the complete reference genome (Ensembl GRCh37.69).

Figure~\ref{fig:Genomes_All} displays the precision and accuracy of Cufflinks, Trinity, and SPATA for the data set consisting of $15$ million reads. Since Trinity did not use any reference genome, its constant performance was used as a benchmark to compare the performance of Cufflinks and SPATA. In Figure~\ref{fig:Genomes_All}, the \emph{left} panel displays the precision of the assembly tools whereas the \emph{right} panel displays their accuracy. The number of reference transcripts reconstructed by Cufflinks, Trinity, and SPATA is displayed in Figure~\ref{fig:Genomes_Counts} in the Appendix.

In terms of precision, the performance of Cufflinks and SPATA are comparable and are consistently higher than that of Trinity. This translates into a robust precision of Cufflinks and SPATA, regardless of quality of the reference genome being used. In terms of accuracy, as the quality of the reference genome declines, the performance of Cufflinks and SPATA also declines and a sharper drop is observed for Cufflinks. For the reference genome sequence hg10, a low-quality reference genome, the accuracy of Trinity and SPATA is almost the same and is higher than that of Cufflinks. For all other reference genome sequences, SPATA achieves a better accuracy than Cufflinks and Trinity.

%\section{Real-world Data Analysis}

\section{Conclusion}

In this paper, we presented SPATA, a hybrid transcriptome assembly approach, to reconstruct transcriptomes via a novel three-stage algorithm: localization, seeding and growing, and patching and cutting. We first assessed the performance of SPATA by reconstructing the transcritomes using multiple simulated RNA-Seq data sets for which SPATA achieves high precision and accuracy. We also showed that SPATA reconstructs more reference transcripts than the selected transcriptome assembly tools for all of the simulated data sets. We further demonstrated the robustness of SPATA against incomplete reference genomes by using different versions of human reference genome over the past $12$ years. SPATA consistently achieved a high precision and reconstructed more reference transcripts than the competing methods. Overall, the analyses favor the use of SPATA in transcriptome reconstruction wherever a reference genome is available, regardless of quality.

SPATA is also expected to be a valuable tool to assemble transcriptomes of non-model organisms, where the reference genome sequences are likely to be incomplete. Despite advances in sequencing technologies, assembling a complete reference genome is still costly and difficult for many species. As a result, most of the existing genomes are available only as unfinished drafts with gaps and excessive assembly errors~\cite{SalzbergSteven2005,SalzbergSteven2012,BakerMonya2012}. In these cases, SPATA is expected to reliably capture both known transcript structures and novel variations due to its high precision and accuracy. SPATA is also conveniently accessible for both informatics and life science researchers via an easy-to-use GUI software.

%\appendices
\section{Appendix} \label{sec:appendix}
\begin{figure}
\centering
\includegraphics[width=0.49\textwidth]{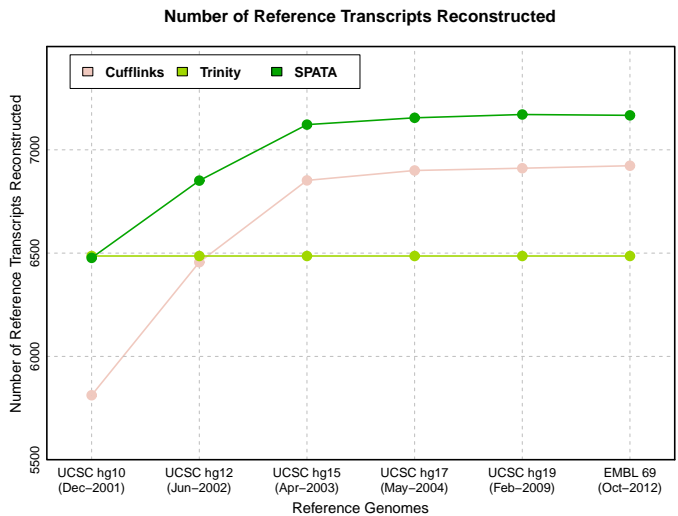}
\caption{The number of reference transcripts reconstructed by Cufflinks, Trinity, and SPATA using different reference genomes for the data set consisting of $15$ million reads. The horizonal axis displays the reference genomes being used in chronological order. The vertical axis displays the total number of reference transcripts that were fully reconstructed. A transcript is consider to be fully reconstructed by a method if it has at least $90\%$ base-identity with a contig from the method's output. We examine all reference transcripts having mean coverage of $5$ or higher.}
\label{fig:Genomes_Counts}
\end{figure}

%\subsection{Simulation Data and Results}
Figure~\ref{fig:Genomes_Counts} displays the number of reference transcripts that were fully reconstructed by Cufflinks, Trinity, and SPATA using different reference genomes for the data set consisting of $15$ million reads. Since the total number of reference transcripts is a constant, the number of the reference transcripts reconstructed by the assembly tools also reflects their accuracy. For the low-quality reference genome genome hg10, SPATA and Trinity reconstruct almost the same number of reference transcripts and reconstruct more than Cufflinks. For all other reference genome sequences, SPATA reconstructs more reference transcripts than Cufflinks and Trinity.

\begin{figure}[h]
\centering
\includegraphics[width=0.49\textwidth]{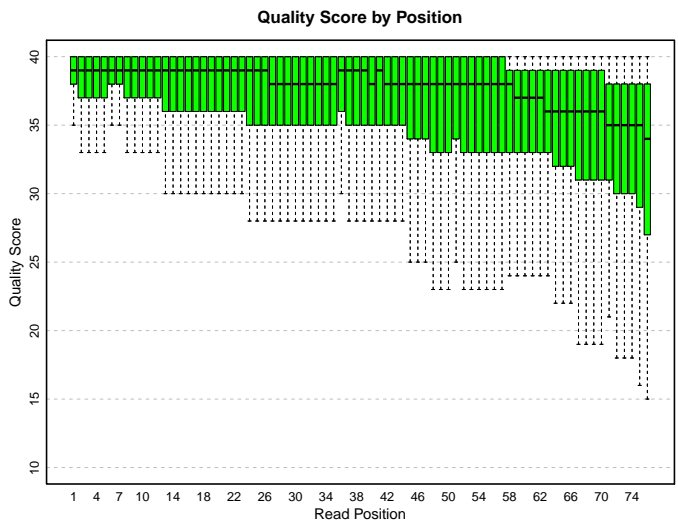}
\caption{Per-base sequencing quality of the simulated data sets. The horizonal axis shows the read position from the $1^{st}$ to the $76^{th}$ position whereas the vertical axis shows the quality scores, which range from $0$ to $40$ (Phred+33). The box plots display the per-base quality distributions at each read position.}
\label{fig:QualityScore}
\end{figure}

\begin{figure}[h!]
\centering
\includegraphics[width=0.49\textwidth]{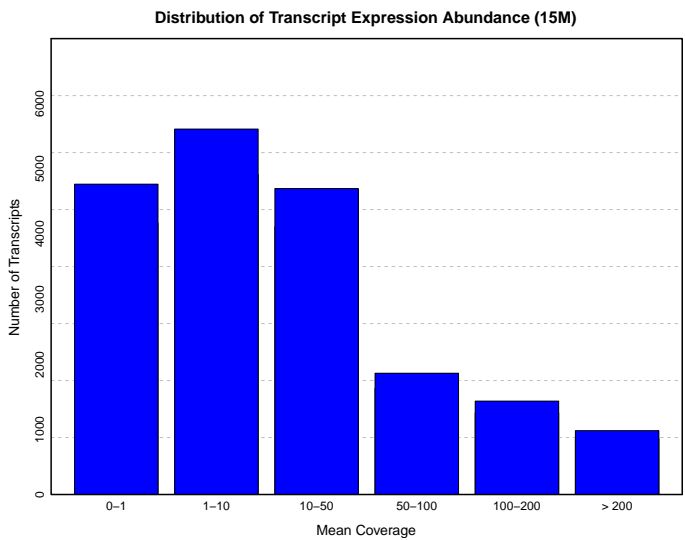}
\caption{The abundance distribution of reference transcripts in simulation study. The horizonal axis represents the mean coverage whereas the vertical axis illustrates the frequency. The data set has $15$ million reads with length $76$.}
\label{fig:AbundanceDistribution}
\end{figure}

The statistics of the simulated data are displayed in Table~\ref{tab:ReferenceTranscripts}, Figure~\ref{fig:QualityScore} and Figure~\ref{fig:AbundanceDistribution}. Table~\ref{tab:ReferenceTranscripts} shows the number of reference transcripts in each data set. Figure~\ref{fig:QualityScore} displays the quality score distribution of sequenced bases across $76$ read positions whereas Figure~\ref{fig:QualityScore} displays the expression abundance distribution of the reference transcripts.

\begin{table}[h]
\centering \caption{Number of reference transcripts}
\begin{scriptsize}
\begin{tabular}{| c | c | c |}
\hline
Data sets&Expressed transcripts&Transcripts having\\
&& coverage $\geq 5$\\
&& \\
\hline
15M&19,351&10,784\\
20M&19,390&11,357\\
30M&19,299&12,326\\
50M&19,316&13,360\\
75M&19,380&14,424\\
100M&19,344&15,160\\
\hline
\end{tabular}
\label{tab:ReferenceTranscripts}
\end{scriptsize}
\end{table}

\bibliographystyle{nar_max_authors_10}
\bibliography{references}

\end{document}